\documentclass[letterpaper, 10 pt, conference]{ieeeconf}  

\IEEEoverridecommandlockouts                              
\usepackage{amsmath,amssymb,comment}

\usepackage{url}
\usepackage{graphicx}
\usepackage{subfigure}
\usepackage{color}
\usepackage{epsfig}
\usepackage{latexsym}

\usepackage{mathtools}

\usepackage{amsmath,amssymb,amsfonts}
\usepackage{mathtools}
\usepackage{dsfont} 
\usepackage{makecell}

\newtheorem{theorem}{\bf Theorem}[section]
\newtheorem{definition}[theorem]{Definition}
\newtheorem{lemma}[theorem]{Lemma}

\usepackage{cite}

\allowdisplaybreaks
\def \OO {{O}}
\def \oo {{o}}
\newcommand{\pr}{\mathbb{P}}

\newcommand{\N}{\mathbb{N}}

\newcommand{\comp}{^{\rm c}}

\newcommand{\hh}{\mathbb{H}(n;K)} 
\newcommand{\hhn}{\mathbb{H}(n;K_n)} 
\newcommand{\hhdn}{\mathbb{H}(n;K_n,\gamma_n)} 

\newcommand{\nodes}{\mathcal{N}}

\title{\LARGE \bf
On the Connectivity and Giant Component Size of Random K-out Graphs Under Randomly Deleted Nodes
}

\author{Eray Can Elumar \and Mansi Sood \and Osman Ya\u{g}an 
\thanks{E.C. Elumar, M. Sood and O. Ya\u{g}an are with Department
of Electrical and Computer Engineering and CyLab,
Carnegie Mellon University, Pittsburgh,
PA, 15213 USA. Email:
{\tt\small \{eelumar@andrew.cmu.edu, msood@andrew.cmu.edu, oyagan@ece.cmu.edu\}}}}

\begin{document}

\maketitle
\thispagestyle{empty}
\pagestyle{empty}




\begin{abstract}
Random K-out graphs, denoted \(\mathbb{H}(n;K)\), are generated by each of the \(n\) nodes drawing \(K\) out-edges towards \(K\) distinct nodes selected uniformly at random, and then ignoring the orientation of the arcs. Recently, random K-out graphs have been used in applications as diverse as random (pairwise) key predistribution in ad-hoc networks, anonymous message routing in crypto-currency networks, and differentially-private federated averaging. In many applications, connectivity  of the random K-out graph when some of its nodes are {\em dishonest}, have {\em failed}, or have been {\em captured} is of  practical interest. We provide a comprehensive set of results on the connectivity and giant component size of \(\mathbb{H}(n;K_n,\gamma_n)\), i.e., random K-out graph when \(\gamma_n\) of its nodes, selected uniformly at random, are deleted. First, we derive conditions for \(K_n\) and \(n\) that ensure, with high probability (whp), the connectivity of the remaining graph when the number of deleted nodes is \(\gamma_n=\Omega(n)\) and \(\gamma_n=o(n)\), respectively. Next, we derive conditions for \(\mathbb{H}(n;K_n,\gamma_n)\) to have a {\em giant component}, i.e., a connected subgraph with \(\Omega(n)\) nodes, whp. This is also done for different scalings of \(\gamma_n\) and  {\em upper} bounds are provided for the number of nodes {\em outside} the giant component. Simulation results are presented to validate the usefulness of the results in the finite node regime.
\end{abstract}


\section{Introduction}
\label{sec:introduction}

Random graphs are widely used in modeling and analysis of diverse real-world networks including social networks~\cite{newman2002random}, economic networks~\cite{kakade2005economic}, and  communication networks~\cite{goldenberg2010survey}. In recent years, a random graph  model known as the {\em random K-out graph} has received interest in designing secure  sensor networks \cite{Yagan2013Pairwise}, decentralized learning  \cite{2020dprivacy}, and anonymity preserving crypto-currency networks \cite{FantiDandelion2018}. Random K-out graphs, denoted $\hh$, are generated over a set of $n$ nodes as follows. Each of the $n$ nodes draws $K$ out-edges towards $K$ distinct nodes selected uniformly at random. The resulting {\em undirected} graph obtained by ignoring the orientation of the edges is referred to as a random K-out graph.

In the context of sensor networks, random K-out graphs have been used \cite{Yagan2013Pairwise, yagan2012modeling, yavuz2015designing} to  analyze  the performance of the random {\em pairwise} key predistribution scheme \cite{Haowen_2003} and its heterogeneous variants \cite{eletreby2020connectivity,sood2020size}. The random {\em pairwise} scheme works as follows. Before deployment, each sensor chooses $K$ others uniformly at random. A unique {\em pairwise} key is given to each node pair where at least one of them selected the other. After deployment, two sensors can securely communicate if they share a pairwise key. The topology of the sensor network can thus be represented by a random K-out graph; each edge of the random K-out represents a secure communication link between two sensors.
Consequently, random K-out graphs have been analyzed  to answer key questions on the values of the parameters $n, K$ needed to achieve certain desired properties, including connectivity at the time of deployment \cite{FennerFrieze1982,Yagan2013Pairwise}, connectivity under {\em link} removals  \cite{yagan2012modeling,yavuz2015designing}, and unassailability \cite{yagan2016wireless}.

Despite many prior works on random K-out graphs, very little is known about its connectivity properties when some of its {\em nodes} are removed. This is an increasingly relevant problem since many deployments of sensor networks are expected to take place in {\em hostile} environments where nodes may be captured by an adversary, or fail due to harsh conditions. In addition, random K-out graphs have  recently been used to construct the communication graph in a differentially-private federated averaging scheme called the GOPA (GOssip Noise for Private Averaging) protocol \cite[Algorithm~1]{2020dprivacy}.  
According to the GOPA protocol, a random K-out graph is constructed on a set of nodes of which an unknown subset is {\em dishonest}. 
It was shown in \cite[Theorem~3]{2020dprivacy} that the privacy-utility trade-offs achieved by the GOPA protocol is tightly dependent on the subgraph on {\em honest} nodes being {\em connected}. When the subgraph on honest nodes is not connected, it was shown that the performance of GOPA is tied to the {\em size} of the connected components of the honest nodes.

With these motivations in mind, this paper aims to fill a gap in the literature and
 provide a comprehensive set of results on the connectivity and size of the giant component of the random K-out graph when some of its nodes are {\em dishonest}, have {\em failed}, or have been {\em captured}. Let  \(\mathbb{H}(n;K_n,\gamma_n)\) denote the random K-out graph when \(\gamma_n\) of its nodes, selected uniformly at random, are deleted.  First, we provide a set of conditions for $K_n$ and $n$ that ensure, {\em with high probability} (whp), the connectivity of the remaining graph when the number $\gamma_n$ of deleted nodes is  $\Omega(n)$ and $o(n)$, respectively.
 Our result for $\gamma_n = \Omega(n)$ (see Theorem \ref{theorem:thmc_1}) significantly improves a prior result \cite{YAGAN2013493} on the same problem and leads to a {\em sharp} zero-one law for the connectivity of \(\mathbb{H}(n;K_n,\gamma_n)\). Our result for the case $\gamma_n = o({n})$ (see Theorem \ref{theorem:thmc_2}) expands the existing threshold of $K_n \geq 2$ required for connectivity by showing that the graph is still connected whp for $K_n \geq 2$ when $o(\sqrt{n})$ nodes are deleted. We then derive conditions on $K_n$ that leads $\hhdn$ to have a {\em giant component} with an upper bound on the number of nodes allowed outside the giant component. This is also done for both cases $\gamma_n=\Omega(n)$ and $\gamma_n=o(n)$. Finally, we present simulation results when the number of nodes is finite and compare the results with an Erd\H{o}s-R\'enyi graph with same average node degree. 

\section{Notations and the Model}
\label{sec:model}

All random variables are defined on the same probability space $(\Omega, {\mathcal{F}}, \mathbb{P})$ and probabilistic statements are given with respect to the probability measure $\mathbb{P}$. The complement of an event $A$  is denoted by $A\comp$. The cardinality of a discrete set $A$ is denoted by $|A|$. 
All limits are understood with  $n$ going to infinity. If the probability of an event tends to one as $n\rightarrow \infty$, we say that it  occurs with high probability (whp). The statements $a_n = \oo(b_n)$, $a_n=\omega(b_n)$,  $a_n = \OO(b_n)$, $a_n=\Theta(b_n)$, and $a_n = \Omega(b_n)$, used when comparing the asymptotic behavior of sequences $\{a_n\},\{b_n\}$, have their meaning in standard Landau notation. The asymptotic equivalence $a_n \sim b_n$ is used to denote the fact that 
$\lim_{n \to \infty} \frac{a_n}{b_n}=1$.

The random K-out graph is defined on the vertex set $V:=\{v_1, \ldots, v_n\}$ as follows. Let $\nodes:=\{1,2,\dots,n\}$ denote the set vertex labels. For each $i \in \nodes$, let $\Gamma_{n,i} \subseteq \nodes \setminus i$ denote the set of $K_n$ labels corresponding to the nodes selected by $v_i$.
 It is assumed that  $\Gamma_{n,1}, \ldots , \Gamma_{n,n}$ are mutually independent. Distinct nodes $v_i$ and $v_j$ are adjacent, denoted by $v_i \sim v_j$ if at least one of them picks the other. Namely, 
\vspace{-1mm}
\begin{align}
v_i \sim v_j ~~\quad \mbox{if} ~~~\quad [j \in \Gamma_{n,i}] ~\vee~ [i \in \Gamma_{n,j}]. 
\label{eq:Adjacency}
\end{align}
The random graph defined on the vertex set $V$ through the adjacency condition (\ref{eq:Adjacency}) is called a random K-out graph \cite{frieze2016introduction,Bollobas,Yagan2013Pairwise} 
and denoted by $\hhn$. 
It was previously established in  \cite{Yagan2013Pairwise, FennerFrieze1982} that random K-out graphs are connected whp when $K \geq 2$ and not connected when $K=1$; i.e.,
 \vspace{-1mm}
\begin{equation} 
\lim_{n \to \infty} \mathbb{P}\left[ \mathbb{H}(n;K) \text{ is connected}\right] =
\begin{cases}
1 & \mathrm{if} \quad K\geq 2, \\
0 & \mathrm{if} \quad K=1.
\end{cases}
\label{eq:homogeneous_zero_one_law}
\end{equation} 

Next, we model random K-out graphs under random removal of  nodes. As already mentioned, our motivation is to understand the properties of the underlying network when some nodes
are {\em dishonest}, or have {\em failed}, or have been {\em captured}. 
 We let $\gamma_n$ denote the number of such nodes and assume, for simplicity, that they are selected uniformly at random among all nodes in $V$. The case where the set of dishonest/captured/failed nodes are selected carefully by an adversary might also be of interest, but is beyond of the scope of the current paper; see \cite{yagan2016wireless} for partial results in that case. A related model of interest is the random K-out graph under randomly deleted {\em edges}. The connectivity and $k$-connectivity under that case have been studied in \cite{yagan2012modeling,yavuz2015toward,yavuz2017k}.
 
 Formally, let $D\subset V$ denote the set of deleted nodes with  $|D| = \gamma_n$.
 We are interested in the random graph $\hhdn$, defined on the vertex set $R = V \setminus  D$ such that  distinct vertices $v_i$ and $v_j$ (both in $R$) are adjacent if they were  adjacent in $\hhn$; i.e., if  $[j \in \Gamma_{n,i}] ~\vee~ [i \in \Gamma_{n,j}]$.


\begin{definition}[Connected Components]
{\sl
A pair of nodes in a graph $\mathbb{G}$ are said to be {\em connected} if there exists a path of edges connecting them. 
A {\em component} $C_i$ of $\mathbb{G}$ is a subgraph in which any two vertices are connected to each other, and no vertex is connected to a node outside of $C_i$. }
%
\label{def:concomp}
\end{definition}{}

A graph  with $n$ nodes is said to have a {\em giant} component if its largest connected component is of size  $\Omega(n)$.





\section{Main Results and Discussion}
\label{sec:Main Results}
Our main results are presented in Theorems $3.1-3.4$ below. 
Each Theorem addresses a design question as to how we should choose the  parameter $K_n$ 
such that when the given number $\gamma_n$ of nodes are deleted, the remaining graph satisfies the given desired property (e.g., connectivity or a giant component with a specific size) whp.

\subsection{Results on Connectivity}
Let $P(n,K_n,\gamma_n) = \pr \left[ \hhdn \text{ is connected}\right]$.
\begin{theorem}
{\sl 
Let $\gamma_n = \alpha n$ with $\alpha$ in $(0,1)$, and consider a scaling $K:\mathbb{N}_0 \to \mathbb{N}_0$ such that with $c>0$ we
have 
\begin{align}
K_n \sim c \cdot r_1(\alpha, n), \ \  \textrm{where} \quad r_1(\alpha, n) = \frac{\log n}{1 - \alpha - \log \alpha}
\label{eq:threshold_1}
\end{align}
is the threshold function. Then, we have
\begin{align}
& \lim_{n \to \infty} P(n,K_n,\gamma_n) =  \begin{cases}
    1, & \mathrm{if}\quad c > 1\\
    0, & \mathrm{if}\quad 0< c < 1.
  \end{cases}
\label{eq:c_1}
\end{align}
}  \label{theorem:thmc_1}
\end{theorem}
The proof of the {\em one-law} in (\ref{eq:c_1}), i.e., that 
$\lim_{n \to \infty} P(n,K_n,\gamma_n)=1$
if $c>1$, is given in Section \ref{sec:proof}. The {\em zero-law} of (\ref{eq:c_1}),
i.e., that 
$\lim_{n \to \infty} P(n,K_n,\gamma_n)=0$
if $c<1$,
was  established previously in \cite[Corollary 3.3]{YAGAN2013493}. There,
a one-law was also provided: 
under
(\ref{eq:threshold_1}), it was shown that
$\lim_{n \to \infty} P(n,K_n,\gamma_n)$
if $c>\frac{1}{1-\alpha}$, leaving a gap between the thresholds of the zero-law and the one-law. 
Theorem \ref{theorem:thmc_1} presented here fills this gap by establishing a tighter one-law, and constitutes a {\em sharp} zero-one law; e.g., when $\alpha=0.5$, 
 the one-law in \cite{YAGAN2013493} is given with $c>2$, while we show that it suffices to 
have $c>1$. 
\vspace{1mm}
\begin{theorem}
{\sl  Consider a scaling $K:\mathbb{N}_0 \to \mathbb{N}_0$. \\
a) If $\gamma_n = o(\sqrt{n})$, then we have
\begin{align}
\lim_{n \to \infty} P(n,K_n,\gamma_n) = 
    1, \quad \mathrm{if}\quad K_n \geq 2 \ \ \forall n
\label{eq:c_2}
\end{align}
b) If $\gamma_n = \Omega(\sqrt{n})$ and $\gamma_n = o(n)$, and if for some sequence $w_n$, it holds that
$$
K_n = r_2(\gamma_n) + \omega_n, \ \ \textrm{where} \quad 
r_2(\gamma_n) = \frac{\log (\gamma_n)}{\log 2 + 1/2}$$
is the threshold function, then we have
\begin{align}
\lim_{n \to \infty} P(n,K_n,\gamma_n) = 
    1, \quad \mathrm{if}\quad \lim_{n \to \infty}\omega_n = \infty
\label{eq:c_2}
\end{align}
}  \label{theorem:thmc_2}
\end{theorem}
\vspace{-2mm}
Random K-out graph was known \cite{FennerFrieze1982,Yagan2013Pairwise} to be connected whp when $K_n \geq 2$ (viz.~(\ref{eq:homogeneous_zero_one_law})). Theorem \ref{theorem:thmc_2} extends this result by showing that  $K_n \geq 2$ is sufficient to have the random K-out graph remain connected whp  when $o(\sqrt{n})$ of its nodes (selected randomly) are  deleted.

\subsection{Results on the Size of the Giant Component}

Let $C_{max}(n,K_n,\gamma_n)$ denote the set of nodes in the {\em largest}  connected component  of  $\hhdn$ 
and let $P_G(n,K_n,\gamma_n,\lambda_n) :=\mathbb{P}[|C_{max}(n,K_n,\gamma_n)| > n - \gamma_n- \lambda_n]$. Namely, 
$P_G(n,K_n,\gamma_n,\lambda_n)$ is the probability that less than $\lambda_n$ nodes are {\em outside} the largest  component of $\hhdn$.



\begin{theorem}
{\sl 
Let $\gamma_n = o(n)$ and $\lambda_n = \Omega(\sqrt{n})$. Consider a scaling $K:\mathbb{N}_0 \to \mathbb{N}_0$ and let
$$r_3(\gamma_n,\lambda_n) = 1 +  \frac{\log(1+\gamma_n / \lambda_n)}{\log 2 + 1/2}$$
be the threshold function. Then, we have
\begin{align*}
& \lim_{n \to \infty}  P_G(n,K_n,\gamma_n,\lambda_n) =  1, \quad \mathrm{if}\quad K_n > r_3(\gamma_n, \lambda_n),  \ \ \forall n.
\end{align*}
}  \label{theorem:thmg_3}
\end{theorem}
\vspace{-3mm}
We remark that  if $\lambda_n=\beta n$ with $\beta>0$, then  $r_3(\gamma_n,\lambda_n) = 1+o(1)$.
This shows that when $\gamma_n=o(n)$, it suffices to have $K_n \geq 2$ for $\hhdn$ to have a giant component containing $\Omega((1-\beta)n)$  nodes for arbitrary $\beta>0$. 

\vspace{2mm}
\begin{theorem}
{\sl 
Let $\gamma_n = \alpha n$ with $\alpha$ in  $(0,1)$,  $\lambda_n=o(n)$, and $\lambda_n=\omega(1)$. Consider a scaling $K:\mathbb{N}_0 \to \mathbb{N}_0$ and let
$$r_4(\alpha, \lambda_n) = 1 +  \frac{\log(1 + \frac{n \alpha}{\lambda_n}) +\alpha + \log(1-\alpha)}{\frac{1-\alpha}{2} - \log\left(\frac{1+\alpha}{2}\right)}$$
be the threshold function. Then, we have
\begin{align*}
& \lim_{n \to \infty}  P_G(n,K_n,\alpha,\lambda) =  1, \quad \mathrm{if}\quad K_n > r_4(\alpha, x_n), \ \ \forall n.
\end{align*}
}  \label{theorem:thmg_5}
\end{theorem}

\vspace{-3mm}
 Due to space limitations, we only provide a proof of Theorem \ref{theorem:thmc_1} here. 
Proofs of all results are available in 
\cite{icc2021proof}. 

\subsection{Simulation Results}

\begin{figure}[!t]
\centering
\includegraphics[scale=0.275]{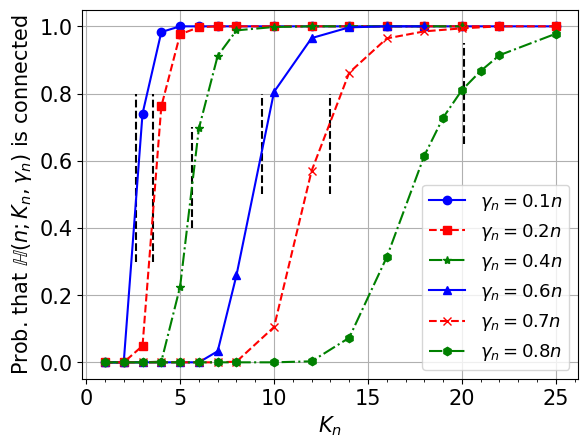}\label{fig:conn1}
\hspace{0.4mm}
\includegraphics[scale=0.275]{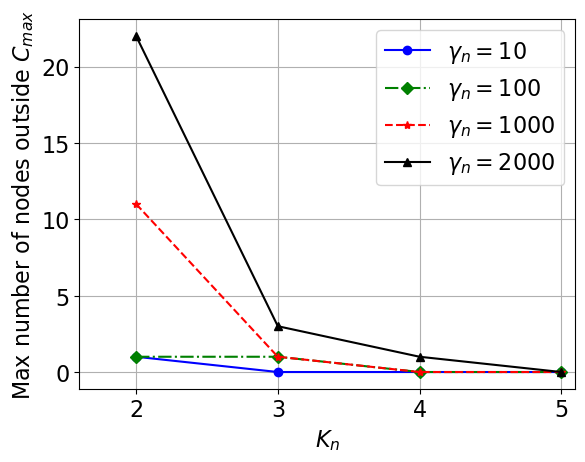}\label{fig:conn2} \vspace{-4mm}
\caption{\sl (Left) Empirical probability that $\hhdn$ is connected for $n = 5000$ calculated from 1000 experiments. The vertical lines are the theoretical thresholds given by Theorem  \ref{theorem:thmc_1}. (Right) Maximum number of nodes outside the giant component of $\hhdn$ for $n = 50,000$ in 1000 experiments. 
\vspace{-4mm}} 
\label{fig:fig1}
\end{figure}

\begin{figure}[!t]
\vspace{-1mm}
\centering
\includegraphics[scale=0.26]{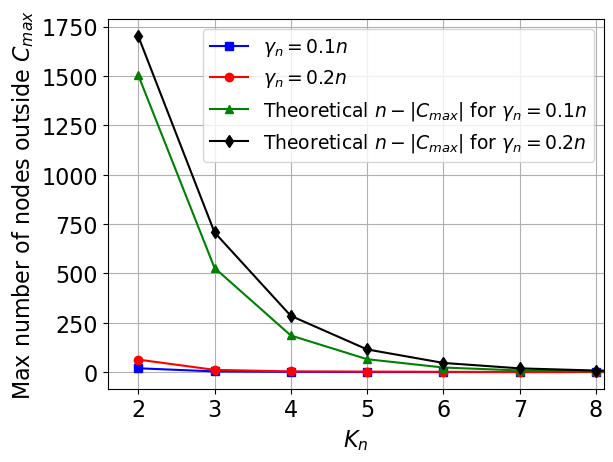}\label{fig:gc2}
\hspace{0.4mm}
\includegraphics[scale=0.27]{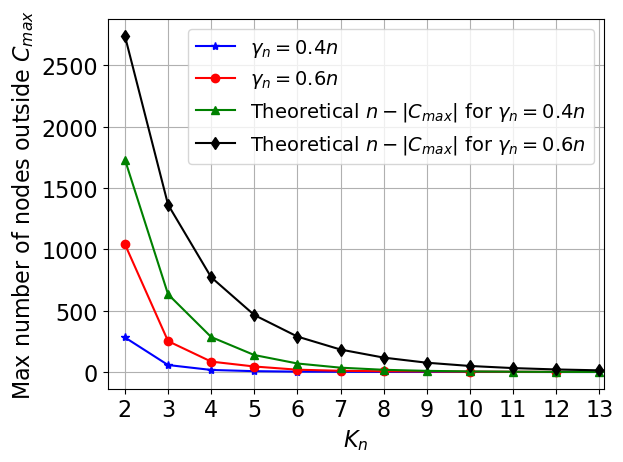}\label{fig:gc2a}  \vspace{-7mm}
\caption{\sl Maximum number of nodes outside the giant component of $\hhdn$ for $n = 5000$ and $\gamma_n = 0.1n$, $\gamma_n = 0.2n$ cases (Left); and for $n = 5000$ and $\gamma_n = 0.4n$, $\gamma_n = 0.6n$ cases (Right), obtained through 1000 experiments along with the respective plot of theoretical $n - |C_{max}|$.} 
\label{fig:fig3}
\end{figure}

To check the usefulness of our results when the number $n$ of nodes is finite, we examine the probability of connectivity and the number of nodes outside the giant component (i.e., $ n - \gamma_n-|C_{max}(n,K_n,\gamma_n)|$) in two different experimental setups. The first setup is to obtain the results for the case where $\gamma_n = \alpha n$, with $\alpha$ in $(0,1)$. We generate  instantiations of the random graph $\hhdn$ with $n=5000$, varying $K_n$  in the interval $[1, 25]$ and several $\alpha$ values in the interval $[0.1,0.8]$. Then, we record the empirical probability of connectivity and $\lambda_n$ from  1000 independent experiments for each $(K_n,\alpha)$ pair. The results of this experiment are shown in Fig.~\ref{fig:fig1} (Left) and Fig.~\ref{fig:fig3}.

Fig.~\ref{fig:fig1} (Left) depicts the empirical probability of connectivity of $\hhdn$. The vertical lines stand for the critical threshold of connectivity asserted by Theorem \ref{theorem:thmc_1}. In each curve, $P(n,K_n,\gamma_n)$ exhibits a threshold behaviour as $K_n$ increases, and the transition from $P(n,K_n,\gamma_n)=0$ to $P(n,K_n,\gamma_n)=1$ takes place around $K_n = \frac{\log n}{1 - \alpha - \log \alpha}$, validating the claims of Theorem \ref{theorem:thmc_1}.

In Fig.~\ref{fig:fig3}, we plot the {\em maximum} number of nodes  outside the giant component observed in 1000 experiments for each parameter pair, and compare these with our result, namely the upper bound on $n - |C_{max}|$ obtained from Theorem \ref{theorem:thmg_5} by taking the maximum $\gamma_n$ value that gives a threshold less than or equal to the $K_n$ value tested in the simulation. As can be seen, for any $K_n$ and $\gamma_n$ value, the experimental maximum number of nodes outside the giant component is smaller than the upper bound obtained from Theorem \ref{theorem:thmg_5}, reinforcing the usefulness of our results in practical settings.


\begin{figure}[!t]
\vspace{-3mm}
\centering
\includegraphics[scale=0.275]{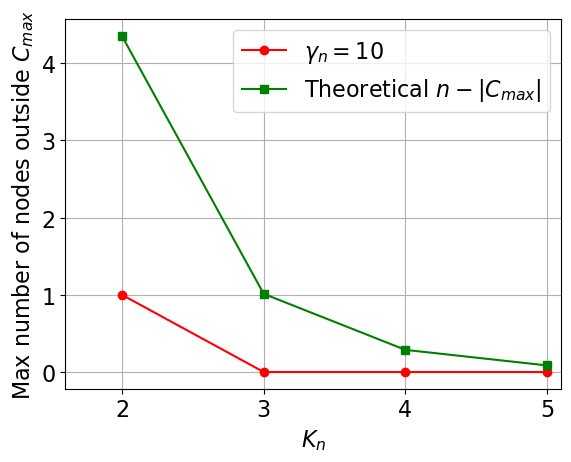}\label{fig:gc1} 
\hspace{0.4mm}
\includegraphics[scale=0.275]{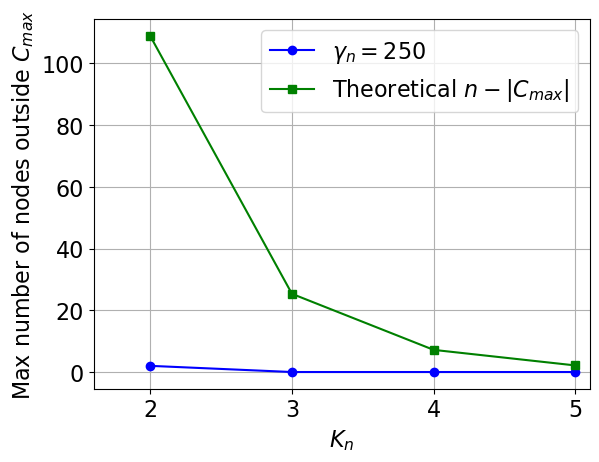}\label{fig:gc3} \vspace{-4mm}
\caption{\sl Maximum number of nodes outside the giant component of $\hhdn$ for $n = 50,000$ and $\gamma_n = 10$ cases (Left); and for for $n = 50,000$ and $\gamma_n = 250$ cases (Right), obtained through 1000 experiments along with the plot of theoretical $n - |C_{max}|$. \vspace{-2mm}} 
\label{fig:fig2}
\end{figure}

\begin{figure}[!t]
\vspace{-1mm}
\centering
\includegraphics[scale=0.26]{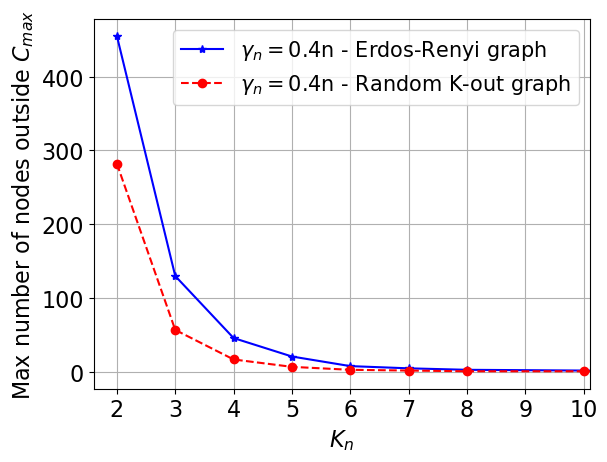}\label{fig:cmp1}
\hspace{0.4mm}
\includegraphics[scale=0.26]{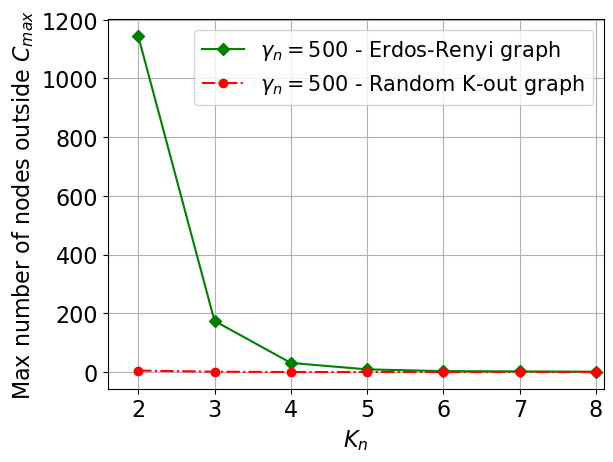}\label{fig:cmp2}  \vspace{-4mm}
\caption{\sl Comparison of maximum number of nodes outside the giant component of a random K-out graph $\hhdn$ and an Erd\H{o}s-R\'enyi graph with same mean node degree when $n = 5000$, $\gamma_n = 0.4n$ (Left); and when $n = 50,000$ and $\gamma_n= 500$ (Right). Each data-point is obtained through 1000 experiments. \vspace{-7mm}} 
\label{fig:cmp}
\end{figure}

We ran a second set of experiments  for the case where $\gamma_n =  o(n)$. As before, we generate  instantiations of the random graph  $\hhdn$,
 with $n=50,000$, varying $K_n$ in  $[2, 5]$ and varying $\lambda_n$ in $[10,2000]$. For each $(K_n,\gamma_n)$ pair, we generate 1000 experiments and record the maximum number of nodes seen outside the giant component; in some case no nodes are seen outside the giant component indicating that the graph is connected. The results of this experiment are shown in Fig.~\ref{fig:fig1} (Right) and Fig.~\ref{fig:fig2}.

In Fig.~\ref{fig:fig1} (Right), the maximum number of nodes seen outside the giant component in 1000 experiments is depicted as a function of $K_n$. The plots for $\gamma_n=10$ and $\gamma_n=100$ correspond to the $\gamma_n = o(\sqrt{n})$ case in Theorem \ref{theorem:thmc_2}a. As can be seen from these plots, there is only one node outside the giant component in the worst case for $\gamma_n=10$ and $\gamma_n=100$ when $K_n = 2$, roughly in line with Theorem \ref{theorem:thmc_2}a which expects the graph to be connected when $K_n = 2$. The plots for $\gamma_n=1000$ and $\gamma_n=2000$ correspond to the $\gamma_n = w(\sqrt{n})$ and $\gamma_n = o(n)$ case in Theorem \ref{theorem:thmc_2}b. The thresholds on $K_n$ for these $\gamma_n$ values, obtained using Theorem \ref{theorem:thmc_2}b are $r_2(1000)=6.79$ and $r_2(2000)=7.37$, rounded to two digits after decimal when the $\omega(1)$ term in Theorem \ref{theorem:thmc_2}b is ignored due to $n$ having a finite value in the simulations. As can be seen from the plots, the graph becomes connected for $\gamma_n=1000$ when $K_n \geq4$, and for $\gamma_n=2000$ when $K_n \geq5$. Hence, we can see that graphs for $\gamma_n=1000$ and $\gamma_n=2000$ are connected when $K_n$ is selected above the theoretical threshold obtained from \ref{theorem:thmc_2}b, supporting Theorem \ref{theorem:thmc_2}b.

In Fig.~\ref{fig:fig2}, the maximum number of nodes seen outside the giant component in 1000 experiments is plotted as a function of $K_n$ for $\gamma_n = 10$ (Left) and for $\gamma_n = 250$  (Right). The corresponding theoretical plots are obtained by the upper bound on $n - |C_{max}|$ asserted by Theorem \ref{theorem:thmg_3} for the given value of $K_n$.
For any $K_n$ and $\gamma_n$ pair, the experimental values are smaller than the theoretical values, supporting the usefulness of Theorem \ref{theorem:thmg_3} in the finite node regime.

\subsection{Discussion}

 In Theorem \ref{theorem:thmc_1}, we improve the results given in \cite{YAGAN2013493} by closing the gap between the zero law and the one law, and hence we establish a sharp zero-one law for connectivity when $\gamma_n= \Omega(n)$ nodes are deleted from $\hhdn$. 
 
 In Theorem \ref{theorem:thmc_2}, we establish that the graph $\hhdn$ with $\gamma_n = o(n)$ is connected whp when $K_n \sim \log(\gamma_n)$; and when $\gamma_n = o(\sqrt{n})$, $K_n \geq 2$ is sufficient for connectivity. The latter result is especially important, since $K_n \geq 2$ is the previously established threshold for connectivity \cite{FennerFrieze1982}, and here we improve this result by showing that the graph is still connected with $K_n \geq 2$ even after $o(\sqrt{n})$ nodes (selected randomly) are  deleted. 

To put these results in perspective, we compare them with an Erd\H{o}s-R\'enyi graph $G(n,p)$, which is connected whp if $p > \log n / n$. This translates to having an average node degree of $<k> \sim \log n$ \cite{erdHos1960evolution}. The $<k>$ required for the random K-out graph to be connected whp is much lower, with $<k> = O(1)$ when $o(\sqrt{n})$ nodes are removed, and $<k> \sim \log(\gamma_n)$ when $\gamma_n=\Omega(\sqrt{n})$ nodes are removed. 

For a better comparison, we examine the experimental maximum number of nodes outside the giant component out of 1000 experiments of a random K-out graph $\hhdn$ and an Erd\H{o}s-R\'enyi graph $G(n,p)$ with same mean node degree when $\gamma_n$ random nodes are removed from the graph. To achieve the same node degree, $p$ is selected as $p = 2K_n/n$. The results are given in Fig.~\ref{fig:cmp} for $n=5000$, $\gamma_n =0.4n$ on (Left), and $n=50,000$, $\gamma_n= 500$ on (Right). As can be seen, the random K-out graph has less maximum number of nodes outside the giant component than the Erd\H{o}s-R\'enyi graph and this difference is more pronounced when $\gamma_n$ is smaller. Hence, we can conclude that random K-out graphs are more robust to random node removals than Erd\H{o}s-R\'enyi graphs in the sense of probability of connectivity and size of the giant component being larger. This reinforces the efficiency of the K-out construction in various distributed network applications including federated averaging \cite{2018dprivacy,2020dprivacy} where it is desirable to maintain connectivity in the event of node failures or adversarial capture of nodes.


\section{A Proof of Theorem~\ref{theorem:thmc_1} }
\label{sec:proof}
\vspace{-1mm}
We start  by  defining a {\em cut}.

\begin{definition}[Cut]


\cite[Definition 6.3]{MeiPanconesiRadhakrishnan2008}
For a graph $\mathcal{G}$ defined on the node set $V$, a \emph{cut} is a non-empty subset $S \subset V$ of nodes {\em isolated} from the rest of the graph. Namely,  $S \subset V$ is a cut if there is no  edge between $S$ and $S\comp=V \setminus S$.  
\label{def:cut}
\end{definition}{}

Definition~\ref{def:cut} implies that if $S$ is a cut, then so is $S\comp$. Recall from Section \ref{sec:model} that we defined $\hhdn$ as the graph when the set $D$ of nodes is removed from the graph $\hhn$. Namely, the vertex set of $\hhdn$ is given by $R=V \setminus  D$.   Let $\mathcal{E}_n (K_n, \gamma_n; S)$ denote the event that  $S \subset R$ is a cut in $\hhdn$ as per Definition~\ref{def:cut}. With $S\comp = R/S$, the event $\mathcal{E}_n (K_n, \gamma_n; S)$ occurs if no nodes in $S$ pick neighbors in $S\comp$, and no nodes in $S\comp$ pick neighbors in $S$. Note that nodes in $S$ or $S\comp$ can still pick neighbors in the set $D$. Thus, we have
\begin{align}
\mathcal{E}_n (K_n, \gamma_n; S) =
\bigcap_{i \in \nodes_S} \bigcap_{j \in \nodes_{S\comp}}
\left(
\left \{ i \not \in \Gamma_{n-\gamma_n,j} \right \}
\cap 
\left \{ j \notin \Gamma_{n-\gamma_n,i} \right \}
\right). \nonumber
\vspace{-2mm}
\end{align}
with $\nodes_S$,  $\nodes_{S\comp}$ denoting the set of labels of the vertices in  $S$ and $S\comp$, respectively.

Let $\mathcal{Z}(x_n;K_n, \gamma_n)$ denote the event that $\hhdn$ has no cut $S \subset R $ with size  $x_n \leq |S| \leq n-\gamma_n - x_n$ where  $x:\N_0 \rightarrow  \N_0$ is a sequence such that $x_n \leq (n-\gamma_n)/{2} \ \forall n$. 
Namely, $\mathcal{Z}(x_n;K_n, \gamma_n)$ is the event that there are no cuts in $\hhdn$ whose size falls in the range $[x_n, n-\gamma_n-x_n]$.

\begin{lemma}
\cite[Lemma 4.3]{sood2020size} For any sequence $x: \mathbb{N}_0 \rightarrow \mathbb{N}_0$ such that $x_n \leq \lfloor (n-\gamma_n)/3 \rfloor$ for all $n$, we have
\begin{align}
    \mathcal{Z}(x_n;K_n, \gamma_n) \Rightarrow |C_{max}(n, K_n, \gamma_n)| > n - \gamma_n - x_n.
\end{align}
\label{lemma:gc}
\end{lemma}
\vspace{-5mm}
Lemma \ref{lemma:gc} states that if the event $\mathcal{Z}(x_n;K_n, \gamma_n)$  holds, then the size of the largest connected component of $\hhdn$ is greater than $n - \gamma_n - x_n$; i.e.,  there are less than $x_n$ nodes outside of the giant component of $\hhdn$. Hence, we can see that $\hhdn$ is connected if $\mathcal{Z}(x_n;K_n, \gamma_n)$ takes place with $x_n=1$. Thus, the one-law will be established if we show that $\lim_{n \to \infty} \pr [\mathcal{Z}(x_n;K_n, \gamma_n)\comp] = 0$ with $x_n=1$. 
From the definition of $\mathcal{Z}(x_n;K_n, \gamma_n)$, we have
\vspace{-1mm}
\begin{align}
\mathcal{Z}(x_n;K_n, \gamma_n) & = \bigcap_{S \in \mathcal{P}_n: ~x_n\leq  |S| \leq \lfloor \frac{n-\gamma}{2} \rfloor}  \left(\mathcal{E}_n({K}_n,{\gamma_n}; S)\right)\comp, \nonumber
\end{align}
where $\mathcal{P}_n$ is the collection of all non-empty  subsets of $R$. Complementing both sides and using union bound, we get
\begin{align}
\pr\left[\left(\mathcal{Z}(x_n;K_n, \gamma_n)\right)\comp\right] &\leq \hspace{-3mm}  \sum_{ S \in \mathcal{P}_n: x_n \leq |S| \leq \lfloor \frac{n-\gamma}{2} \rfloor } \pr[ \mathcal{E}_n ({K}_n,{\gamma_n}; S) ] \nonumber \\
&=\hspace{-1mm} \sum_{r=x_n}^{ \left\lfloor \frac{n-\gamma}{2} \right\rfloor } \hspace{-1mm}
 \sum_{S \in \mathcal{P}_{n,r} } \pr[\mathcal{E}_n ({K}_n,{\gamma_n}; S)] \label{eq:BasicIdea+UnionBound},
\end{align}
where  $\mathcal{P}_{n,r} $ denotes the collection of all subsets of $R$ with exactly $r$ elements.
For each $r=1, \ldots , \left\lfloor (n-\gamma)/2\right\rfloor$, we can simplify the notation by denoting $\mathcal{E}_{n,r} ({K}_n,{\gamma_n})=\mathcal{E}_n ({K}_n,{\gamma_n} ; \{v_1, \ldots , v_r \} )$. From the exchangeability of the node labels and associated random variables, we have
\[
\pr[ \mathcal{E}_n({K}_n,{\gamma_n} ; S) ] = \pr[ \mathcal{E}_{n,r}({K}_n,{\gamma_n}) ], \quad S \in
\mathcal{P}_{n,r}.
\]
$|\mathcal{P}_{n,r} | = {n-\gamma_n \choose r}$, since there are ${n-\gamma_n \choose r}$ subsets of $R$ with r elements. Thus, we have
\begin{equation*}
\sum_{S \in \mathcal{P}_{n,r} } \pr[\mathcal{E}_n ({K}_n,{\gamma_n} ; S) ] 
= {n-\gamma_n\choose r} ~ \pr[\mathcal{E}_{n,r} ({K}_n,{\gamma_n})]. 
\label{eq:ForEach=r}
\end{equation*}
Substituting this into (\ref{eq:BasicIdea+UnionBound}), we obtain 
\begin{align}
\hspace{-.4mm} \pr\left[\left(\mathcal{Z}(x_n;K_n,\gamma_n)\right)\comp\right] \leq \hspace{-1mm} \sum_{r=x_n}^{ \left\lfloor \frac{\hspace{-.5mm} n-\gamma}{2} \right\rfloor } \hspace{-1mm} 
{n-\gamma_n \hspace{-.4mm} \choose r } \hspace{-.4mm}  \pr[ \mathcal{E}_{n,r}({K}_n,{\gamma_n})] \hspace{-1mm} 
\label{eq:Z_bound}
\end{align}
\vspace{-0.1mm}
Remember that $\mathcal{E}_{n,r}({K}_n,{\gamma_n})$ is the event that the $n-\gamma_n-r$ nodes in $S$ and $r$ nodes in $S\comp$ do not pick each other; but they can pick from the $\gamma_n$ nodes from $D$. Thus, we have
 \begin{align} \nonumber
\pr [\mathcal{E}_{n,r}({K}_n,{\gamma_n})] & =   \left( \dfrac{{\gamma_n+r-1 \choose K_n}}{{n-1 \choose K_n}} \right)^{r} \left( \dfrac{{n-r-1 \choose K_n}}{{n-1 \choose K_n}} \right)^{n-\gamma_n-r} \\
 & \leq \left(\dfrac{\gamma_n+r}{n}\right)^{rK_n} 
\left(\dfrac{n-r}{n}\right)^{K_n(n-\gamma_n-r)} 
\nonumber
\end{align}
Letting $P_Z=\pr\left[\mathcal{Z}(1;K_n, \gamma_n)\comp\right]$, and plugging in $x_n=1$ in (\ref{eq:Z_bound}), we get 
\begin{align}
\hspace{-1mm} P_Z \leq \hspace{-1mm}  \sum_{r=1}^{ \left\lfloor \frac{n-\gamma}{2} \right\rfloor } \hspace{-1mm} {\hspace{-.5mm}n-\gamma_n\hspace{-.5mm}\choose r} \hspace{-1mm} \left(\hspace{-.5mm}\dfrac{\hspace{-.5mm}\gamma_n+r\hspace{-.5mm}}{n}\right)^{\hspace{-1mm} r K_n} 
\hspace{-1mm} \left(\hspace{-.5mm}\dfrac{n-r}{n}\hspace{-.5mm} \right)^{\hspace{-1mm} K_n(n-\gamma_n-r)\hspace{-.5mm}} 
\label{eq:gc_pz}
\end{align}
Let  $\gamma_n = \alpha n$ with $0<\alpha<1$. Using this and standard bounds
${n \choose k} \leq (\frac{n e }{k})^k$ and $1-x \leq e^{-x}$
in  (\ref{eq:gc_pz}), we get
\begin{align}
P_Z \leq \sum_{r=1}^{ \left\lfloor \frac{n-\alpha n}{2} \right\rfloor } \left(\dfrac{n -\alpha n}{r}\right)^r e^r \left(\alpha + \dfrac{r}{n}\right)^{rK_n} e^{\frac{-r K_n (n-\alpha n -r)}{n}} \nonumber
\end{align}
We will show that the right side of the above expression goes to zero as $n$ goes to infinity. Let
 $$A_{n,r,\alpha}: = \left(\dfrac{n -\alpha n}{r}\right)^r e^r \left(\alpha + \dfrac{r}{n}\right)^{rK_n}e^{\frac{-rK_n(n-\alpha n -r)}{n}}.$$
 \vspace{-2mm}
We write
 \vspace{-1mm}
\begin{align*}
P_Z \leq \sum_{r=1}^{ \left\lfloor n/\log n \right\rfloor } A_{n,r,\alpha} + \sum_{r=\left\lfloor n/\log n \right\rfloor}^{ \left\lfloor \frac{n-\alpha n}{2} \right\rfloor } A_{n,r,\alpha} := Q_1 + Q_2,
\end{align*}
and show that both $Q_1$ and $Q_2$ go to zero as $n \to \infty$.
We start with the first summation $Q_1$.
 \vspace{-1mm}
\begin{align}
Q_1 & \! \begin{multlined}[t]
 \leq \sum_{r=1}^{ \left\lfloor \frac{n}{\log n}  \right\rfloor } \left((1 -\alpha )en\cdot e^{  K_n \log(\alpha+ \frac{1}{\log n})  -K_n (1-\alpha -\frac{1}{\log n}) } \right)^r \nonumber
 \end{multlined}
  \vspace{-1mm}
 \end{align}
 \vspace{-1mm}
Next, assume as in the statement of Theorem \ref{theorem:thmc_1} that 
\begin{align}
    K_n = \frac{c_n \log n}{1 - \alpha - \log \alpha}, \quad n=1,2,\ldots
\label{eq:proof1_k}
\end{align}
 for some sequence $c: \mathbb{N}_0 \to \mathbb{R}_+$ such that $\lim_{n \to \infty}c_n = c$ with $c>1$. 
Also define
\begin{align*}
a_n &:=  (1 -\alpha )en\cdot e^{  K_n\log(\alpha+ \frac{1}{\log n})  -K_n(1-\alpha -\frac{1}{\log n}) }  \\
& = (1 -\alpha )e n \cdot e^{- c_n \log n \frac{1- \alpha -\log (\alpha+\frac{1}{\log n}) - \frac{1}{\log n}}{1- \alpha - \log \alpha}}
\\
& = (1 -\alpha )e n^{1-c_n} e^{c_n \frac{\log n \log(1+ \frac{1}{\alpha \log n}) +1}{1- \alpha - \log \alpha}}
\\
& = O(1) n^{1-c_n}
\end{align*}
where we substituted $K_n$ via (\ref{eq:proof1_k}) and used the fact that $ \log n \cdot \log(1+ \frac{1}{\alpha \log n}) \leq   \frac{1}{\alpha}$. Taking the limit as $n \to \infty$ and recalling that 
$\lim _{n \to \infty} c_n = c >1$, we see that
$\lim _{n \to \infty} a_n = 0$. Hence, for large $n$, we  have
\begin{align}
Q_1 \leq \sum_{r=1}^{ \left\lfloor n/\log n  \right\rfloor } \left( a_n \right)^r \leq \sum_{r=1}^{ \infty} \left( a_n \right)^r = \frac{a_n}{1-a_n}
\end{align}
where the geometric sum converges by virtue of $\lim _{n \to \infty} a_n = 0$. Using this once again, it is clear from the last expression that 
$\lim _{n \to \infty}Q_1 = 0$.

Now, similarly consider the second summation $S_2$.
\begin{align}
 Q_2 \leq \sum_{r=  \lfloor n/\log n\rfloor }^{ \left  \lfloor (n-\alpha n)/2 \right\rfloor } \left((1 - \alpha )e\log n \cdot e^{ K_n \log(\frac{1+\alpha}{2}) -K_n\frac{1- \alpha}{2} } \right)^r  \nonumber
\end{align}
\vspace{-2mm}
Next, we define 
\begin{align}
b_n &:= (1 - \alpha )e\log n \cdot e^{ -K_n \left(  \frac{1-\alpha}{2}  - \log(\frac{1+\alpha}{2}) \right)}
\end{align}
Substituting for $K_n$ via (\ref{eq:proof1_k}) and taking the limit as ${n \to \infty}$ it can be seen that
$
\lim _{n \to \infty} b_n = 0
$
upon noting that $ \frac{1-\alpha}{2}  - \log(\frac{1+\alpha}{2}) >0$ and $\lim _{n \to \infty} c_n =c >1$.
 With  arguments similar to those used in the case of $Q_1$, we can show that when $n$ is large $S_2 \leq b_n/(1-b_n)$, leading to $Q_2$ converging to zero as $n$ gets large.
With $P_Z \leq Q_1 + Q_2$, and both $Q_1$ and $Q_2$ converging to zero when $n$ is large, we establish that $P_Z$ converges to zero as $n$ goes to infinity. This result also yields the desired conclusion $\lim _{n \to \infty} P(n,K_n,\gamma_n) =1$ in Theorem \ref{theorem:thmc_1} since $P_Z = 1-P(n,K_n,\gamma_n)$. We direct readers to \cite{icc2021proof} for proof of other Theorems presented in Section \ref{sec:Main Results}. 
\vspace{-4mm}
\section{Conclusions}
\vspace{-1mm}
In this paper, we provide a comprehensive set of results on the connectivity and giant component size of $\hhdn$, i.e.,  random K-out graph with randomly selected $\gamma_n$ nodes deleted. Computer simulations are used to validate the results in the finite node regime. Using our results, we compare  random K-out graphs with  Erd\H{o}s-R\'enyi graphs with the same mean node degree and same number of deleted nodes, and show that random K-out graphs are either connected with higher probability or have a larger giant component. This reinforces the usefulness of random K-out graphs in various distributed network applications including federated averaging \cite{2018dprivacy,2020dprivacy}.  Our results can help design networks with desired levels of robustness or tolerance to  nodes failing, being captured, or being dishonest  in many applications including wireless sensor networks and distributed averaging.
\vspace{-1mm}
\section*{Acknowledgements}
This work was supported by the National Science Foundation through Grant \# CCF-1617934, and by CyLab through the Secure and Private IoT Initiative.

  \bibliographystyle{IEEEtran}
\bibliography{IEEEabrv,references}

\end{document}